\documentclass[final,times,11p,twocolumn]{elsarticle}
\usepackage{amssymb}
\usepackage{amsmath}
\usepackage{lineno}
\usepackage{listings}

\usepackage{color}

\usepackage{hyperref}
\hypersetup{colorlinks, linkcolor=blue}

\makeatletter
\def\ps@pprintTitle{%
  \let\@oddhead\@empty
  \let\@evenhead\@empty
  \let\@oddfoot\@empty
  \let\@evenfoot\@oddfoot
}
\makeatother

\begin{document}

\definecolor{codegreen}{rgb}{0,0.6,0}
\definecolor{codegray}{rgb}{0.5,0.5,0.5}
\definecolor{codepurple}{rgb}{0.58,0,0.82}
\definecolor{backcolour}{rgb}{0.95,0.95,0.92}
 
\lstdefinestyle{mystyle}{
    commentstyle=\color{codegreen},
    keywordstyle=\color{magenta},
    numberstyle=\tiny\color{codegray},
    stringstyle=\color{codepurple},
    basicstyle=\ttfamily\footnotesize,
    breakatwhitespace=false,         
    breaklines=true,                 
    captionpos=b,                    
    keepspaces=true,                 
    numbers=left,                    
    numbersep=5pt,                  
    showspaces=false,                
    showstringspaces=false,
    showtabs=false,                  
    tabsize=2
}
 
\lstset{style=mystyle}

\sloppy
\begin{frontmatter}

\title{A generalized nonlinear Schrödinger Python module implementing different models of input pulse quantum noise}

\author[add1,add2]{O. Melchert\corref{mycorrespondingauthor}}
\ead{melchert@iqo.uni-hannover.de}
\cortext[mycorrespondingauthor]{Corresponding author}

\author[add1,add2]{A. Demircan}
\ead{demircan@iqo.uni-hannover.de}

\address[add1]{Leibniz Universit\"at Hannover, Institute of Quantum Optics (IQO), 30167 Hannover, Germany}
\address[add2]{Cluster of Excellence PhoenixD (Photonics, Optics, and Engineering - Innovation Across Disciplines), Hannover, Germany}

\begin{abstract}
We provide Python tools enabling numerical simulation and analysis of the propagation dynamics of ultrashort laser pulses in nonlinear waveguides.  
The modeling approach is based on the widely used generalized nonlinear Schrödinger equation for the pulse envelope.
The presented software implements the effects of linear dispersion, pulse self-steepening, and the Raman effect.
The focus lies on the implementation of input pulse shot noise, i.e.\ classical background fields that mimick quantum noise, which are often not thoroughly presented in the scientific literature.
We discuss and implement commonly adopted quantum noise models based on pure spectral phase noise, as well as Gaussian noise.
Coherence properties of the resulting spectra can be calculated.
We demonstrate the functionality of the software by reproducing results for a supercontinuum generation process in a photonic crystal fiber, documented in the scientific literature.
The presented Python tools are are open-source and released under the MIT license in a publicly available software repository.
\end{abstract}

\begin{keyword}
Generalized nonlinear Schrödinger equation \sep quantum noise \sep spectral coherence \sep Python
\end{keyword}

\end{frontmatter}

\tableofcontents

\section{Introduction}

The propagation of laser pulses in nonlinear waveguides supports the generation of supercontinuum spectra \cite{Agrawal:BOOK:2019,Mitschke:BOOK:2016,Dudley:RMP:2009}. 
Starting from a spectrally narrow input pulse, the interplay of linear and nonlinear effects induces tremendous spectral broadening, yielding flat spectra that can extend from the violet to the infrared \cite{Ranka:OL:2000}. 
Such effects can be achieved, e.g., in photonic crystal fibers (PCFs) \cite{Knight:N:2003,Birks:OL:1997}, wherein supercontinuum spectra can be produced using $\sim\!100\,\mathrm{fs}$-duration pulses, peak powers $\sim\!10\,\mathrm{kW}$ and propagation lengths on the order of $1\,\mathrm{m}$ \cite{Ranka:OL:2000}. 
The resulting broad, flat spectra with high spectral density find application, e.g., in optical frequency metrology \cite{Jones:S:2000}, and optical technologies \cite{Mitschke:BOOK:2016}.

A flexible theoretical framework for
studying the complex physical processes associated with the generation of supercontinuum spectra is provided by the generalized nonlinear Schrödinger equation (GNLS) \cite{Agrawal:BOOK:2019}.
In order to model the propagation dynamics of laser pulses it combines the effects of linear dispersion, pulse self-steepening \cite{deMartini:PR:1967,deOliveira:JOSAB:1992}, and the Raman effect \cite{Gordon:OL:1986}.
This accounts for various processes that support the generation of widely extended supercontinuum spectra, 
such as the modulation istability \cite{Demircan:OC:2005}, soliton-fission \cite{Husakou:PRL:2001,Demircan:APB:2007}, and self-frequency shift of Raman solitons \cite{Gordon:OL:1986}. 
It furthers forms the basis for modeling optical rouge waves \cite{Solli:N:2007,Solli:PRL:2008,Demircan:SR:2012,Dudley:NP:2014}, and the interaction of solitons and dispersive waves in presence of an optical event-horizon \cite{Demircan:PRL:2011,Demircan:OE:2014,Skryabin:RMP:2010,Melchert:NCP:2020}.
The GNLS describes the real-valued optical field in terms of a complex-valued envelope at a fixed reference frequency using a nonlinear wave equation of first order in the propagation coordinate. It has proven to yield reliable results even in the single cycle regime \cite{Brabec:PRL:1997}.
Overall, the GNLS provides a classical description of the propagation dynamics of input pulses. 
Basic phenomena described by the GNLS are susceptible to the presence of noise.
When attempting to model noise on the quantum level, a 
fundamental issue arises since the rigorous description of quantum noise sources requires the introduction of quantum mechanical operators \cite{Haus:JOSAB:1990,Henry:RMP:1996,Drummond:JOSAB:2001,Corney:JOSAB:2001}, which are incompatible with classical fields.
When the number of photons in the input pulse is large, however, semiclassical approaches exist that account for quantum fluctuations by randomizing physical interactions using a classical background field of low power, entering the GNLS via the initial condition.  
Such input pulse noise is modeled by simply adding to the coherent input pulse an incoherent, stochastic noise field.
We consider commonly adopted one photon per frequency mode 
\cite{Dudley:RMP:2009,Dudley:JSTQ:2002,Dudley:OL:2002,Kobtsev:OE:2005,Genier:JOSAB:2019},
and half a photon per temporal mode \cite{Paschotta:APB:2004,Ruehl:PRA:2011,Zhou:APL:2016} noise models, and further implement a background field based on a classical analog of the zero-point field of quantum field theory (QFT) \cite{Ibson:PRA:1996}.
Including noise by perturbing the initial conditions is in accord with the Wigner method, an efficient computational approach for modeling quantum noise in optical fibers \cite{Corney:JOSAB:2001,Drummond:JOSAB:2001,Werner:PRA:1995}.
%
%
Considering an ensemble of independent simulation runs perturbed by noise, coherence properties of the simulated spectra can be investigated.
Accounting for shot-to-shot fluctuations in numerical simulations
appears essential for the dynamics and enables a better comparison to actual experiments, employing multishot measurement techniques \cite{Dudley:JSTQ:2002,Dudley:OL:2002}.

In this article we present software tools allowing to reliably simulate and analyze basic phenomena described by the GNLS in presence of various types of input pulse shot noise. 
We detail the scientific problem addressed by the provided software in Sec.~\ref{sec:model}, and discuss its implementation, with emphasis on input pulse noise models, in Sec.~\ref{sec:description}.
Section \ref{sect:sampleResults} reports verification tests, reproducing results for a supercontinuum generation process in a photonic crystal fiber, well documented in the scientific literature, where it has been used for illustrating nonlinear-optics effects \cite{Dudley:RMP:2009}, and for benchmarking algorithms \cite{Hult:JLT:2007,Heidt:JLT:2009,Rieznik:IEEEPJ:2012,Balac:CPC:2016,Melchert:CPC:2022}.
We comment on impact and conclude in Sec.~\ref{sec:conclusions}.

\section{Scientific problem solved by the software \label{sec:model}}


%

The provided Python tools enable simulation and analysis of the dynamics of ultrashort laser pulses, governed by the generalized nonlinear Schrödinger equation (GNLS) \cite{Agrawal:BOOK:2019,Dudley:RMP:2009}
\begin{align}
\partial_z u = i \sum_{n\geq 2}& \frac{\beta_n}{n!}(i\partial_t)^n u + i \gamma \left( 1+\frac{i\partial_t}{\omega_0}\right) \notag\\
&\times \left[\,u(z,t)\int R(t^\prime) |u(z,t-t^\prime)|^2~{\rm{d}}t^\prime\right],
                                   \label{eq:GNLS}
\end{align}
for a complex-valued field envelope $u\equiv u(z,t)$
on a periodic time-domain of extend $T$ with boundary condition $u(z,-T/2)=u(z,T/2)$ and propagation distance $z$. An initial condition $u(z=0,t)=u_0(t)$ needs to be specified. 
%
%
%
In Eq.~(\ref{eq:GNLS}), $\beta_n$ (in units of $\mathrm{ps^n/\mu m}$) is the $n$-th order dispersion coefficient, $\gamma$ ($\mathrm{W^{-1}/\mu m}$) is a scalar nonlinear coefficient, and $\omega_0$ ($\mathrm{rad/ps}$) a reference frequency.
$t$ is a retarded time, measured in a reference frame moving with the group velocity at $\omega_0$.
%
The Raman effect is included via the total response function  
\begin{align}
R(t)=(1-f_{\rm{R}})\,\delta(t) + f_{\rm{R}}\,h_{\rm{R}}(t), \label{eq:R}
\end{align}
where the first term results in an instantaneous Kerr-type nonlinear response, and
where the Raman response function $h_{\rm{R}}(t)$ enters with fractional contribution $f_{\rm{R}}$ \cite{Blow:JQE:1989,Stolen:JOSAB:1989}.
A generic two-parameter model is
\begin{align}
h_{\rm{R}}(t) = \frac{\tau_1^2 + \tau_2^2}{\tau_1 \tau_2^2}\,e^{-t/\tau_2}\,\sin(t/\tau_1)\,\Theta(t), \label{eq:hR_t}
\end{align}
with the Heaviside step function $\Theta(t)$ ensuring causality \cite{Blow:JQE:1989}. 
Based on a single damped harmonic-oscillator approximation of molecular responses, 
this simple model reproduces the Raman gain spectrum measured for silica glass reasonably well.
%
 For silica fibers, numerical values for the three relevant model parameters are $f_{\rm{R}}=0.18$, $\tau_1 = 12.2~\mathrm{fs}$, and $\tau_2=32~\mathrm{fs}$ \cite{Blow:JQE:1989}.
More complex response functions, based on improved response models and valid for other media, exist \cite{Lin:OL:2006,Hollenbeck:JOSAB:2002,Agger:JOSAB:2012}. 
%
In this framework, using a discrete sequence of angular frequency detunings $\Omega\in \frac{2\pi}{T}\mathbb{Z}$ relative to $\omega_0$, the expressions
\begin{subequations}\label{eq:FT}
\begin{align}
&\mathsf{F}[u(z,t)] \equiv \frac{1}{T} \int_{-T/2}^{T/2} u(z,t)\,e^{i\Omega t}~{\rm{d}}t = u_\Omega(z),\label{eq:FT_FT}\\
&\mathsf{F}^{-1}[u_\Omega(z)] \equiv \sum_{\Omega} u_\Omega(z)\,e^{-i\Omega t} =  u(z,t) , \label{eq:FT_IFT}
\end{align}
\end{subequations}
specify forward [Eq.~(\ref{eq:FT_FT})], and inverse [Eq.~(\ref{eq:FT_IFT})] Fourier transforms, relating the field envelope $u(z,t)$ to the spectral envelope $u_\Omega(z)$.
%
Using Parseval's identity for the transforms (\ref{eq:FT}), the energy in both domains is 
\begin{align}
E(z) = \int_{-T/2}^{T/2} |u(z,t)|^2~{\rm{d}}t = T \sum_\Omega |u_\Omega(z)|^2, \label{eq:E}
\end{align}
with instantaneous power $|u(z,t)|^2$ ($\mathrm{W=J/s}$) and power spectrum $|u_\Omega(z)|^2$ ($\mathrm{W}$). 
%
%
While Eq.~(\ref{eq:GNLS}) represents the GNLS in its time-domain formulation, a consistent formulation in the frequency domain is also possible \cite{Francois:JOSAB:1991}.


Input pulse noise is modeled by simply adding to the coherent input pulse $u_0(t)$ an incoherent, stochastic noise field $\Delta u(t)$ with properties
\begin{subequations}\label{eq:noise}
\begin{align}
&\langle u(t) \rangle = 0,\label{eq:noise_mean}\\
&\langle \Delta u(t) \Delta u^{*}(0)\rangle = \sigma^2\delta(t), \label{eq:noise_corr} 
\end{align}
\end{subequations}
where $\langle \ldots\rangle$ denotes an ensemble average over independent instances of noise, and $\sigma^2$ is the noise variance. 
According to Eq.~(\ref{eq:noise_mean}), at each point $t$, this background field has zero mean.
The $\delta$-function in the autocorrelation (\ref{eq:noise_corr}) indicates that the noise varies fast in comparison to any reasonable field $u_0(t)$, and that subsequent actions of the noise are uncorrelated. 
For instance, considering \emph{half a photon} with energy $\hbar \omega_0$ per temporal mode \cite{Paschotta:APB:2004,Ruehl:PRA:2011,Zhou:APL:2016}, the noise variance is $\sigma^2=\hbar \omega_0/2$ with reduced Planck constant $\hbar$.
%
%
Noise models for treating more specific pump laser spectra exist \cite{Frosz:OE:2006,Frosz:OE:2010}.
Let us note that there also exist formulations of the propagation dynamics in terms of stochastic nonlinear Schrödinger equations, which are equivalent to quantum field operator equations and account for vaccuum fluctuations during propagation \cite{Drummond:PRL:1987,Drummond:JOSAB:1987,Brainis:PRA:2005}.

In the frequency domain, shot-to-shot fluctuations of the field can then be characterized by the spectrally resolved modulus of first order coherence for zero time-lag 
\begin{align}
|g_{12}(\Omega)| = \left| \frac{\langle u_{\Omega,m}^{\phantom{*}} u_{\Omega,k}^* \rangle_{m\neq k} }{\sqrt{\langle|u_{\Omega,m}|^2\rangle \langle|u_{\Omega,k}|^2\rangle }}\right|, \label{eq:g}
\end{align}
where the angular brackets specify an average over nonidentical pairs of fields, labeled $m$ and $k$, obtained from an ensemble of $M$ simulation runs with independent noise fields \cite{Dudley:JSTQ:2002,Dudley:OL:2002,Dudley:RMP:2009}. At a given angular frequency detuning $\Omega$, $0 \leq |g_{12}(\Omega)|\leq 1$, where $|g_{12}(\Omega)|\approx 1$ indicates good stability in amplitude and phase.
Different from this measure of interpulse coherence, the modified intrapulse coherence 
\begin{align}
\tilde{\Gamma}(\Omega_1,\Omega_2) = \frac{|\langle u^{\phantom{*}}_{\Omega_1,m} u^{*}_{\Omega_2,m}\rangle|}{\langle |u^{\phantom{*}}_{\Omega_1,m} u^{*}_{\Omega_2,m}| \rangle} \label{eq:IPC}
\end{align}
allows to assess the coherence between different spectral components within a pulse. With focus on $f$-to-$2f$ ($f=$ frequency) setups, the intrapulse coherence was shown to play an important role in carrier-envelope phase measurement and stabilization of ultrashort pulses \cite{Raabe:PRL:2017}. Equation~(\ref{eq:IPC}) modifies the intrapulse coherence of Ref.~\cite{Raabe:PRL:2017} by relaxing the $f/2f$ condition. Instead, two general distinct frequencies $\Omega_1$, and $\Omega_2$ are considered.  

%

\section{Software description\label{sec:description}}

{\tt GNLStools} is written using the Python programming language \cite{Rossum:1995}, and depends on the functionality of numpy, scipy \cite{Virtanen:NM:2020}, and matplotlib \cite{Hunter:CSE:2007}. It can be cloned directly from GitHub \cite{GNLStools:GitHub:2022}, where it is available under the MIT license.

\subsection{Software Functionalities}

The current version of {\tt GNLStools} features:
\begin{itemize}
\item A data structure for the GNLS (\ref{eq:GNLS}), which can be easily tailored to a specific nonlinear waveguide.

\item Functions that implement optional input pulse  noise models considering both, pure spectral phase noise as well as Gaussian noise.

\item Functions for calculating the coherence properties of simulated spectra.
\end{itemize}

The software can be used on its own, it includes a basic driver script implementing a fixed stepsize ``fourth-order Runge Kutta in the interaction picture'' (RK4IP) solver \cite{Hult:JLT:2007} (Sec.\ \ref{sec:results01}), or as extension module for {\tt py-fmas} \cite{Melchert:CPC:2022} (Sec.\ \ref{sec:results02}).
In the latter case, a user can take advantage of variable stepsize $z$-propagtion algorithms and more elaborate Raman-response functions.
The presented software is not overly feature-rich, but allows to reliably simulate the complex physical processes that enable generation of supercontinuum spectra in nonlinear waveguides \cite{Dudley:RMP:2009}.

\subsection{Implementation details}
Solving the GNLS in the time-domain formulation Eq.~(\ref{eq:GNLS}) presents several disadvantages. Higher-oder dispersion is difficult to handle. Even for the lowest order $t$-derivatives, introduction of finite-difference operations comes with unwanted truncation errors.
It is therefore advantageous to perform some of the operations in the frequency-domain, allowing to employ spectral derivatives $\partial_t^n\, u = \mathsf{F}^{-1}[(-i\Omega)^n \,u_\Omega]$.
The operating principles of common algorithms for the solution of nonlinear Schrödinger-type equations, like the split-step Fourier method \cite{Taha:JCP:1984,DeVries:AIP:1987}, the RK4IP method \cite{Hult:JLT:2007}, and the variable step-size conservation quantity error method \cite{Heidt:JLT:2009,Rieznik:IEEEPJ:2012}, exploit this.
To facilitate implementation, the interval $-T/2\ldots T/2$ is divided into $N$ equal subintervals with temporal grid spacing $\Delta t=N/T$, yielding discrete grid points $t_m=m\, \Delta t$, and detuning grid points $\Omega_m=m\,\Delta \Omega$, $\Delta \Omega = 2\pi/T$, with $m=-N/2,\ldots,N/2-1$. Subsequently we write $u(z,t)|_{t=t_m}= u_m(z)$, and $u_\Omega (z)|_{\Omega=\Omega_m} = u_{\Omega_m}(z)$.
Respecting the sign choice and normalization of the transforms (\ref{eq:FT}), we use the numpy native discrete Fourier transform (DFT) routine {\tt ifft} to implement Eq.~(\ref{eq:FT_FT}), and {\tt fft} to implement Eq.~(\ref{eq:FT_IFT}) \cite{note:FT}.


\subsubsection{The GNLS data structure}
Instantiating an instance of the class {\tt GNLS} requires a user to specify several input parameters. Below, they are listed as ``{\tt parameter\_name} (data type): description'':
\begin{itemize}
    \item {\tt w} (array, float): discrete angular frequencies $(\Omega_m)_{m=-N/2}^{N/2-1}$ in units of $\mathrm{rad/fs}$;
    \item {\tt beta\_n} (array, float): ordered sequence of dispersion parameters $(\beta_n)_{n=2}^{N_{\rm{max}}}=(\beta_2, \beta_3, \ldots, \beta_{N_{\rm{max}}})$, with $N_{\rm{max}}\geq 2$ and $\beta_n$ in units of $\mathrm{fs^n/\mu m}$;
    \item {\tt gamma} (float): nonlinear parameter $\gamma$ $(\mathrm{W^{-1}/\mu m})$;
    \item {\tt w0} (float): reference angular frequency $\omega_0$ $(\mathrm{rad/fs})$;
    \item {\tt fR} (float): Raman contribution $f_{\rm{R}}$ (default: $0.18$);
    \item {\tt tau1} (float): time-scale $\tau_1$ $(\mathrm{fs})$ (default: $12.2\,\mathrm{fs}$);
    \item {\tt tau2} (float): time-scale $\tau_2$ $(\mathrm{fs})$ (default: $32\,\mathrm{fs}$).
\end{itemize}

While the angular frequency grid {\tt w} is passed as positional argument, all other parameters are passed as keyword arguments. An example is discussed in Sect.~\ref{sec:results01}.
For using an instance of the {\tt GNLS} class with a $z$-propagation algorithm, several instance methods are available. Below, they are listed in the format ``{\tt method\_name(arg1,arg2,...)}: description'':

\begin{subequations}
\begin{description}
    \item[{\tt Lw(self)}:] 
    Using spectral derivatives, the frequency domain representation of the linear operator
    $L \equiv i \sum_{n\geq 2} \frac{\beta_n}{n!}(i\partial_t)^n$ on the right-hand-side (rhs) of Eq.~(\ref{eq:GNLS}) can be written as
    \begin{align}
    L_\Omega \equiv i\sum_{n= 2}^{N_{\rm{max}}} \frac{\beta_n}{n!} \Omega^n. \label{eq:Lw}
    \end{align}
    For practical reasons, the sum has to be truncated at a finite integer number $N_{\rm{max}}\geq 2$. 
    This class method returns $L_\Omega$, evaluated at the points  $(\Omega_m)_{m=-N/2}^{N/2-1}$.
    
    \begin{description}
    \item {\bf{Input parameters}}:
    \begin{itemize}
    \item The method relies on class attributes only.
    \end{itemize}
    \item {\bf{Output parameters}}:
    \begin{itemize}
    \item {\tt{Lw}} (array): $L_\Omega$ [Eq.~(\ref{eq:Lw})] at the points $(\Omega_m)_{m=-N/2}^{N/2-1}$.
    \end{itemize}
    \end{description}
    The method is decorated as \verb!@property!, so it can be conveniently used as \verb!gnls_instance.Lw!.

    \item[{\tt Nw(self, uw)}:]
    Using the convolution theorem for the transforms (\ref{eq:FT}), 
    the second term on the rhs of Eq.~(\ref{eq:GNLS}) can be written in a mixed representation as
    \begin{align}
    N_\Omega(u) \equiv i \gamma(1+\omega_0^{-1}\Omega)\,\mathsf{F}\left[ (1\!-\!f_{\rm{R}}) \mathcal{I}_1 + f_{\rm{R}} \mathcal{I}_2  \right], \label{eq:Nw}
    \end{align}
    where $\mathcal{I}_1=|u|^2u$, $\mathcal{I}_2 = \mathsf{F}^{-1}[ \tilde{h}_{\rm{R}}(\Omega)\mathsf{F}[|u|^2]] u$, and
    \begin{align}
    \tilde{h}_{\rm{R}}(\Omega) \equiv T\,\mathsf{F}[h_{\rm{R}}(t)] = \frac{\tau_1^{-2} + \tau_2^{-2}}{\tau_1^{-2} - (\Omega+i \tau_2^{-1})^{2}}. \label{eq:hR_w} 
    \end{align}
    The method takes on input the spectral envelope $u_\Omega$, retrieves $u=\mathsf{F}^{-1}[u_\Omega]$, and evaluates and returns Eq.~(\ref{eq:Nw}) at $(\Omega_m)_{m=-N/2}^{N/2-1}$.
    \begin{description}
    \item {\bf{Input parameters}}:
    \begin{itemize}
    \item {\tt uw} (array): Spectral envelope $u_{\Omega_m}$ at the angular frequency grid points $(\Omega_m)_{m=-N/2}^{N/2-1}$.
    \end{itemize}
    \item {\bf{Output parameters}}:
    \begin{itemize}
    \item {\tt{Nw}} (array): $N_\Omega$ [Eq.~(\ref{eq:Nw})] at the points $(\Omega_m)_{m=-N/2}^{N/2-1}$.
    \end{itemize}
    \end{description}

     \item[{\tt claw\_Ph(self,i,zi,w,uw)}:] 
     
      Class method evaluating a conservation law of the GNLS (\ref{eq:GNLS}), related to the classical analog of the photon number \cite{Blow:JQE:1989}.
      This method considers the energy (\ref{eq:E}) in the form $E=\hbar \sum_\Omega n_\Omega \,(\omega_0+\Omega)$, where the dimensionless quantity $n_{\Omega} \equiv T |u_{\Omega}|^2/[\hbar (\omega_0 + \Omega)]$ specifies the number of photons with energy $\hbar (\omega_0\!+\!\Omega)$. The total number of photons is then given by
    \begin{align}
        C_{\rm{Ph}}(z) \equiv \sum_\Omega n_\Omega = \frac{2\pi}{\hbar \Delta \Omega}\sum_\Omega \frac{|u_\Omega(z)|^2}{\omega_0 + \Omega}. \label{eq:CN}
    \end{align}
     In a rigorous quantum mechanical treatment, the photon number is instead defined using photon creation and annihilation operators \cite{Henry:RMP:1996}.
    Let us note that while the GNLS (\ref{eq:GNLS}) conserves the total number of photons Eq.~(\ref{eq:CN}), it does not conserve the pulse energy Eq.~(\ref{eq:E}) \cite{Blow:JQE:1989}.

    \begin{description}
    \item {\bf{Input parameters}}:
    \begin{itemize}
    \item {\tt i} (int): Integer label of the current $z$-propagation step.
    \item {\tt zi} (float): Current propagation distance.
    \item {\tt w} (array): Angular frequency grid $(\Omega_m)_{m=-N/2}^{N/2-1}$.
    \item {\tt uw} (array): Spectral envelope $u_{\Omega_m}$ at the angular frequency grid points $(\Omega_m)_{m=-N/2}^{N/2-1}$.
    \end{itemize}
    \item {\bf{Output parameters}}:
    \begin{itemize}
    \item {\tt{Cph}} (float): Total number of photons $C_{\rm{Ph}}$ at the current point along $z$ [Eq.~\ref{eq:CN}].
    \end{itemize}
    \end{description}
    
     The method has the structure of an user-action function for use with {\tt py-fmas} \cite{Melchert:CPC:2022} (Sec.~\ref{sec:results02}).

\end{description}
\end{subequations}

\subsubsection{Implemented noise models \label{sec:noise_models}}

We provide optional functions for generating time-domain representations of input pulse noise, consistent with Eqs.~(\ref{eq:noise}).
Below, they are listed in the format ``{\tt function\_name(arg1,arg2,...)}: description'':

\begin{description}

\begin{subequations}
\item[{\tt noise\_model\_01(t,w0,s0)}:]
Function generating an instance of noise by directly sampling in the time-domain. The underlying noise model assumes normally distributed  amplitudes.
\begin{description}
\item {\bf{Input parameters}}:
  \begin{itemize}
    \item {\tt t} (array): Time-grid $(t_m)_{m=-N/2}^{N/2-1}$.
    \item {\tt w0} (float): Pulse center frequency $\omega_0$.
    \item {\tt s0} (int): Integer seed $s_0$ for pseudo random number generator.
  \end{itemize}
\item {\bf{Output parameters}}:
  \begin{itemize}
    \item {\tt du\_t} (array): Instance $(\Delta u_m)_{m=-N/2}^{N/2-1}$ of time-domain noise.
  \end{itemize}
\end{description}

An instance of this type of noise is obtained by directly sampling a sequence $(\Delta u_m)_{m=-N/2}^{N/2-1}$ of complex-valued noise-amplitudes in the time domain as
\begin{align}
\Delta u_m &= \sqrt{\frac{\hbar \omega_0}{4 \Delta t}}\,(X+iY), \label{eq:du_t}
\end{align}
where the independent random variables $X$ and $Y$ yield  independent identically distributed (iid) standard normal random numbers [i.e.\ $X,Y \sim \mathcal{N}(0,1)$]. 
%
%
%
Performing an ensemble average, this type of noise exhibits the properties
\begin{align}
 &\langle \Delta u_m  \rangle = 0, \\
 &\langle \Delta u_m \Delta u_0^* \rangle= \frac{\hbar \omega_0}{2\Delta
t}\,\delta_{m 0},\label{eq:noise_ds_ac}
\end{align}
i.e.\ the average field is zero and the autocorrelation (\ref{eq:noise_ds_ac}) is 
equivalent to Eq.~(\ref{eq:noise_corr}) on a discrete grid
with grid spacing $\Delta t$ \cite{Paschotta:APB:2004}.
An instance of this noise has average energy $\langle \sum_m |u_m|^2 \Delta t\rangle = N \hbar \omega_0/2$, corresponding, on average, to \emph{half a photon} with energy $\hbar \omega_0$ per temporal mode.
\end{subequations}

\begin{subequations}
\item[{\tt noise\_model\_02(t,w0,s0)}:]
Function generating an instance of time-domain noise by sampling its Fourier representation. The underlying noise model assumes pure phase noise.
\begin{description}
\item {\bf{Input parameters}}:
  \begin{itemize}
    \item {\tt t} (array): Time-grid $(t_m)_{m=-N/2}^{N/2-1}$.
    \item {\tt w0} (float): Pulse center frequency $\omega_0$.
    \item {\tt s0} (int): Integer seed $s_0$ for pseudo random number generator.
  \end{itemize}
\item {\bf{Output parameters}}:
  \begin{itemize}
    \item {\tt du\_t} (array): Instance $(\Delta u_m)_{m=-N/2}^{N/2-1}$ of time-domain noise.
  \end{itemize}
\end{description}
An instance of this type of noise is obtained using a Fourier method by first sampling a sequence 
$(\Delta u_{\Omega_m})_{m=-N/2}^{N/2-1}$ of random complex-valued spectral amplitudes
\begin{align}
    \Delta u_{\Omega_m} \equiv \sqrt{\frac{\hbar (\omega_0\!+\!\Omega_m)}{T}}  e^{-i\Phi},\label{eq:du_w}
\end{align}
where the random variable $\Phi$ yields iid phase-angles uniformly distributed in the range $0\ldots2\pi$ [i.e.\ $\Phi\sim U(0,2\pi)$];
an inverse Fourier
transform, consistent with Eq.~(\ref{eq:FT_IFT}), is then used to obtain the sequence $(\Delta u_{m})_{m=-N/2}^{N/2-1}$ of time-domain noise amplitudes.

For this type of noise, the magnitude of the spectral amplitudes (\ref{eq:du_w}) is definite and the energy per mode $\Omega$ is $T |u_\Omega|^2=\hbar (\omega_0\!+\!\Omega)$.
With the number of photons $n_\Omega$ as per Eq.~(\ref{eq:CN}), adding noise as $u_{\Omega} \rightarrow u_{\Omega} + \Delta u_{\Omega}$ results in $n_{\Omega} \rightarrow n_{\Omega} + 
1$. 
The photon occupation number is thus increased by the minimal definite amount of \emph{one (entire) photon} per mode.
The phase of this photon is, however, entirely indefinite and the noise in different modes has no particular phase relationship.
An instance of the noise has total energy $T \sum_\Omega |u_\Omega|^2=N \hbar \omega_0$. 

\end{subequations}

\item[{\tt noise\_model\_03(t,w0,s0)}:]
Function generating an instance of time-domain noise by sampling its Fourier representation. The underlying noise model assumes normally distributed spectral amplitudes.
\begin{description}
\item {\bf{Input parameters}}:
  \begin{itemize}
    \item {\tt t} (array): Time-grid $(t_m)_{m=-N/2}^{N/2-1}$.
    \item {\tt w0} (float): Pulse center frequency $\omega_0$.
    \item {\tt s0} (int): Integer seed $s_0$ for pseudo random number generator.
  \end{itemize}
\item {\bf{Output parameters}}:
  \begin{itemize}
    \item {\tt du\_t} (array): Instance $(\Delta u_m)_{m=-N/2}^{N/2-1}$ of time-domain noise.
  \end{itemize}
\end{description}

An instance of this type of noise is obtained using a Fourier method by
first sampling a sequence
$(\Delta u_{\Omega_m})_{m=-N/2}^{N/2-1}$ of random complex-valued spectral amplitudes
\begin{align}
    \Delta u_{\Omega_m} \equiv &\sqrt{\frac{\hbar (\omega_0\!+\!\Omega_m)}{T}} \sqrt{I}  e^{-i\Phi},\label{eq:cnoise}
\end{align}
where the random variable $I$ obeys an exponential distribution with rate parameter $2$  [i.e.\ $I\sim {\mathrm{Exp}}(2)$ with expectation value $\langle I \rangle = 1/2$], and $\Phi$ is uniformly distributed in $0\ldots2\pi$ [i.e.\ $\Phi\sim U(0,2\pi)$];
an inverse Fourier transform, consistent with Eq.~(\ref{eq:FT_IFT}), is then used to obtain the sequence $(\Delta u_{m})_{m=-N/2}^{N/2-1}$ of time-domain noise amplitudes.

Averaging over the noise in a mode $\Omega$ yields energy $\langle T |\Delta u_\Omega|^2 \rangle= \hbar (\omega_0\!+\!\Omega) \langle I \rangle = \hbar (\omega_0\!+\!\Omega)/2$, corresponding, on average, to  \emph{half a photon} per mode.
In contrast to the pure phase noise model implemented by {\tt noise\_model\_02}, this type of noise exhibits instance to instance fluctuations of the energy in each mode. Thereby, the random variable $I$ takes the role of an occupation number with support $I\in[0,\infty)$. 
On average, an instance of this noise has energy $\langle T \sum_\Omega |u_\Omega|^2 \rangle = N \hbar \omega_0/2$. 
%
%
Alternatively, using the Box-M\"uller transform \cite{Box:AMS:1958}, we may rewrite $\sqrt{I} e^{-i \Phi}$, with $I\sim{\rm{Exp}}(2)$, $\Phi \sim U(0,2\pi)$, in Eq.~(\ref{eq:cnoise}) as $(X + iY)/2$, with $X,Y \sim \mathcal{N}(0,1)$ two iid standard normal random variables.
This noise model is based on a classical analog of the zero-point field of QFT, which exactly reproduces the statistics of the electromagnetic vacuum \cite{Ibson:PRA:1996}.

\end{description}

\begin{figure}[t!]
\includegraphics[width=\linewidth]{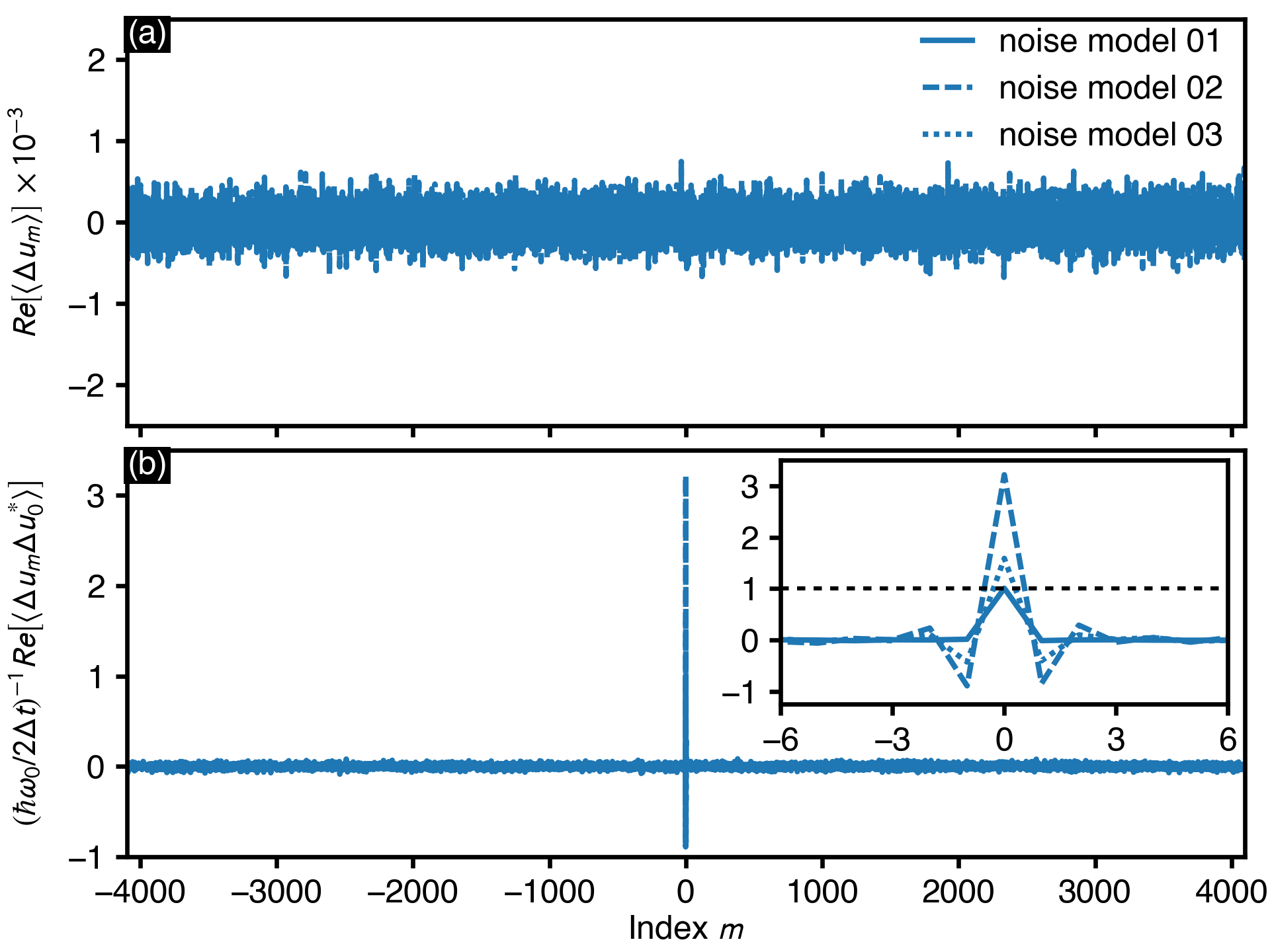}
\caption{Ensemble averaged noise moments.
(a) Real part of the time-domain noise amplitudes as function of the grid point index $m=t_m/\Delta t$. 
(b) Scaled autocorrelation. The inset shows a close up view of the index range $-6 \ldots 6$.
Noise is sampled on a grid with temporal extent $T=4~\mathrm{(ps)}$, $N=2^{13}$ grid points, and for $\omega_0=2.2559~\mathrm{rad/fs}$.
Averages are performed over $10^4$ independent instances of noise.
\label{fig:00}}
\end{figure}

Fourier methods that obtain time-domain noise by sampling frequency dependent spectral noise amplitudes generally result in correlations $\langle \Delta u_m \Delta u^{*}_0 \rangle \sim |t_m|^{-1}$. This is demonstrated in Fig.~\ref{fig:00}, where the ensemble averaged field and autocorrelation for the three noise models are shown. As evident from the inset in Fig.~\ref{fig:00}(b), both Fourier methods result in correlations that persist over many grid spacings $\Delta t$.

Let us note that the above noise models differ with respect to details of implementation and interpretation. Nevertheless, with some degree of approximation, they can easily be related to each other.
For instance, under the assumption that the required bandwidth of the computational domain is small compared to the pulse center frequency, i.e.\ \mbox{$N \Delta \Omega \ll \omega_0$}, we may approximate Eq.~(\ref{eq:du_w}) by $\Delta u_{\Omega_m}\approx\sqrt{\hbar \omega_0/T} \exp(-\Phi)$. All terms in the inverse Fourier transform [Eq.~(\ref{eq:FT_IFT})] of these noise spectral amplitudes are then identically distributed. Approximating these sums using the central limit theorem yields Eq.~(\ref{eq:du_t}) times a factor of $\sqrt{2}$
(this is a consequence of one model assuming one entire photon per mode, while the other assumes half-a photon per mode).
Let us note that this approximation is accompanied by difficulties: it corresponds to \emph{one (entire) photon} per mode if the energy
(\ref{eq:E}) is written as $E=\hbar \omega_0
\sum_\Omega n_\Omega^\prime$, with $n_\Omega^\prime\equiv T |u_\Omega|^2 /(\hbar \omega_0)$. 
Thus, irrespective of $\Omega$, all photons are assumed to contribute the same energy $\hbar \omega_0$. 
Put into the context of pulse propagation models, this is a feature of standard nonlinear Schrödinger-type equations for which energy conservation and photon number conservation are trivially linked.
This, however, is at odds with the GNLS (\ref{eq:GNLS}) which conserves the number of photons but not the energy \cite{Blow:JQE:1989}.
Under the same assumption and approximation, Eq.~(\ref{eq:cnoise}) can easily be related to Eq.~(\ref{eq:du_t}): applying the inverse Fourier transform, we may rewrite the individual terms using the Box-M\"uller transform \cite{Box:AMS:1958} to directly obtain Eq.~(\ref{eq:du_t}).
Moreover, in terms of a Fourier method and in the above approximation, practically any iid spectral amplitudes will yield time-domain noise with properties Eqs.~(\ref{eq:noise}) \cite{Goedecke:FP:1983}.
As a technical detail, let us note that when using a Fourier method to set up noise, the noise spectral amplitudes also depend on the normalization of the transform pair (\ref{eq:FT}). For instance, in Ref.~\cite{Genier:JOSAB:2019}, a GNLS on the infinite $t$-domain with a different transform pair and, consequently, spectral noise amplitudes with different normalization, was considered. 

%
%

\subsubsection{Functions for calculating coherence properties \label{sec:coherence_properties}}

We provide optional functions for calculating the coherence properties of spectra obtained from pulse propagation simulations in terms of the GNLS (\ref{eq:GNLS}).
Below, they are listed in the format ``{\tt function\_name(arg1,arg2,...)}: description'':

\begin{description}
\item[{\tt coherence\_interpulse(w,uw\_list)}:]
Function computing the interpulse coherence Eq.~(\ref{eq:g}).
\begin{description}
\item {\bf{Input parameters}}:
  \begin{itemize}
    \item {\tt w} (array): Angular-frequency grid $(\Omega_m)_{m=-N/2}^{N/2-1}$.
    \item {\tt uw\_list} (array): List \verb![uw1, uw2, ...]!, comprising independent spectra \verb!uw1, uw2, ...!, obtained for the same propagation distance $z$ but for different instances of input pulse noise.
  \end{itemize}
\item {\bf{Output parameters}}:
  \begin{itemize}
    \item {\tt g12} (array):
    $g_{12}(\Omega)$ [Eq.~(\ref{eq:g})] at the points  $(\Omega_m)_{m=-N/2}^{N/2-1}$.
  \end{itemize}
\end{description}

\item[{\tt coherence\_intrapulse(w,uw\_list,w1,w2)}:]
Function computing the intrapulse coherence Eq.~(\ref{eq:IPC}).
\begin{description}
\item {\bf{Input parameters}}:
  \begin{itemize}
    \item {\tt w} (array): Angular-frequency grid $(\Omega_m)_{m=-N/2}^{N/2-1}$.
    \item {\tt uw\_list} (array): List \verb![uw1, uw2, ...]!, comprising independent spectra \verb!uw1, uw2, ...!, obtained for the same propagation distance $z$ but for different instances of input pulse noise.
    \item {\tt w1} (float): Reference angular frequency $\Omega_1$.
    \item {\tt w2} (float): Reference angular frequency $\Omega_2$. 
  \end{itemize}
\item {\bf{Output parameters}}:
  \begin{itemize}
    \item {\tt G} (float):
    $\tilde{\Gamma}^{\rm{CEP}}(\Omega_1,\Omega_2)$ [Eq.~\ref{eq:IPC}].
  \end{itemize}
\end{description}
\end{description}

\begin{figure}[b!]
\includegraphics[width=\linewidth]{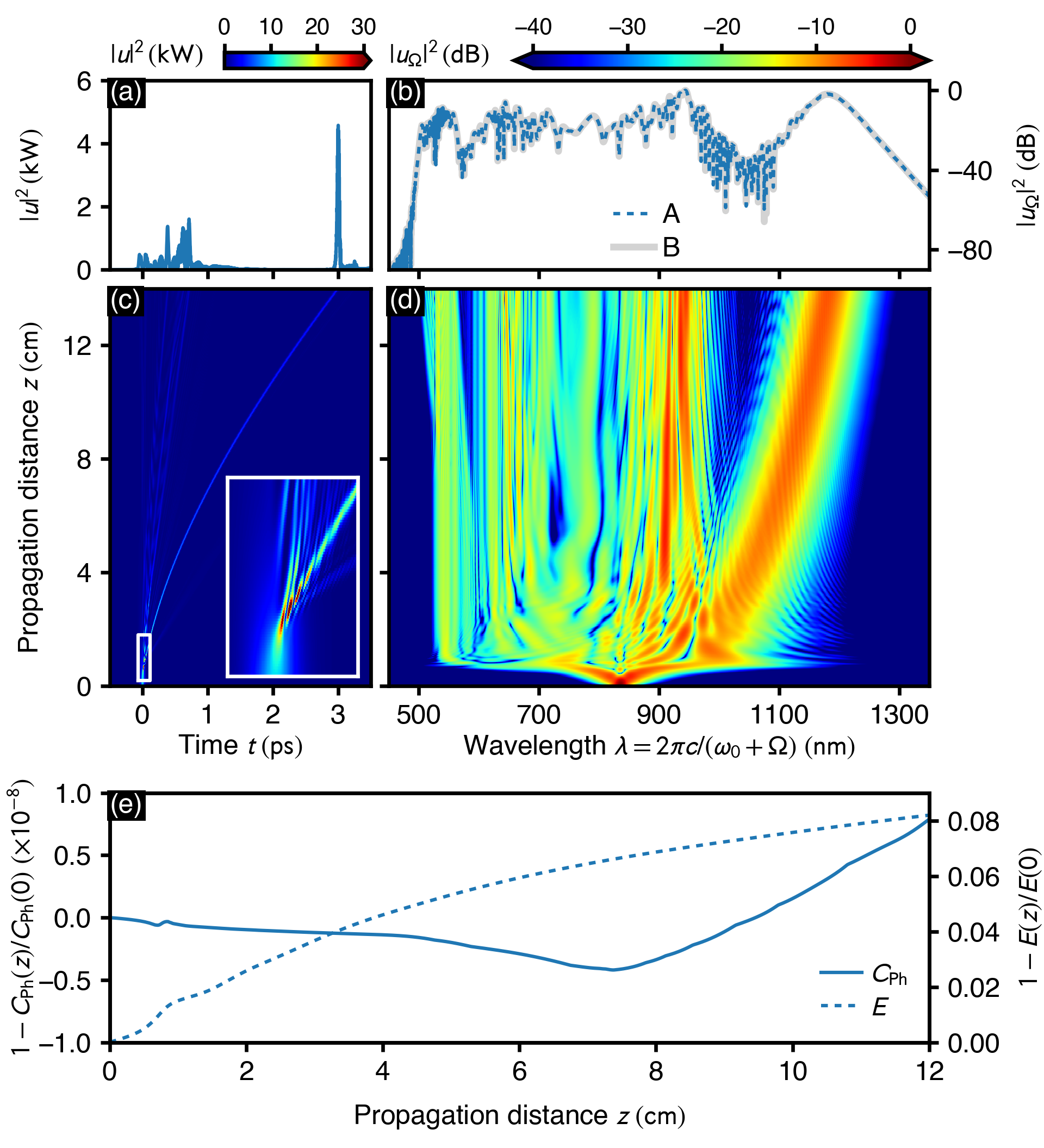}
\caption{Supercontinuum generation in a photonic crystal fiber.
(a) Instantaneous power, and, (b) spectrum after $14\,\mathrm{cm}$ of propagation. In (b), result of the presented implementation (labeled A) is compared to pyNLO \cite{pyNLO:GH:2013} (labeled B).
(c) Propagation of the instantaneous power, and, (d) 
spectrum as function of propagation distance. White boxes in (c) show a close-up view of the soliton fission process.
(d) Normalized deviation of photon number ($C_{\rm{Ph}}$) and energy ($E$) along the fiber.
\label{fig:01}}
\end{figure}

\section{Sample results \label{sect:sampleResults}}

As a verification test of the presented Python tools, we  consider a supercontinuum generation process in a photonic crystal fiber, detailed in Ref.~\cite{Dudley:RMP:2009}.
Specifically, we perform simulations in terms of the GNLS~(\ref{eq:GNLS}), using the sequence of dispersion coefficients
$\beta_2 = -1.183\times 10^{-2}\,\mathrm{fs^2/\mu m}$, 
$\beta_3 = 8.10383\times 10^{-2}\,\mathrm{fs^3/\mu m}$,
$\beta_4 = -9.5205\times 10^{-2}\,\mathrm{fs^4/\mu m}$,
$\beta_5 = 0.20737\,\mathrm{fs^5/\mu m}$, 
$\beta_6 = -0.53943\,\mathrm{fs^6/\mu m}$,
$\beta_7 = 1.3486\,\mathrm{fs^7/\mu m}$,
$\beta_8 = -2.5495\,\mathrm{fs^8/\mu m}$, 
$\beta_9 = 3.0524\,\mathrm{fs^9/\mu m}$,
$\beta_{10} = -1.7140\,\mathrm{fs^{10}/\mu m}$,
and $\gamma = 0.11 \times 10^{-6}\,\mathrm{W^{-1}/\mu m}$.
For the Raman response we use the standard values for silica fibers $f_{\rm{R}}=0.18$, $\tau_1=12.2\,\mathrm{fs}$, and $\tau_2=32\,\mathrm{fs}$ \cite{Blow:JQE:1989}.
As initial condition we take a hyperbolic-secant pulse
$u_0(t)=\sqrt{P_0}\,{\rm{sech}}(t/t_0)$, with duration $t_0=28.4\,\mathrm{fs}$, peak power $P_0=10\,\mathrm{kW}$, and pump wavelength $\lambda_0=835\,\mathrm{nm}$ corresponding to $\omega_0=2\pi c/\lambda_0\approx 2.2559\,\mathrm{rad/fs}$ (with speed of light $c=0.29979\,\mathrm{\mu m/fs}$).
The number of photons [Eq.~(\ref{eq:CN})] in this pulse amounts to $C_{\rm{ph}}\approx 2.4\times 10^{9}$. 
The $z$-propagation dynamics of the above propagation scenario without noise background, followed over $14\,\mathrm{cm}$, is shown in Figs.~\ref{fig:01}(a-d).

\subsection{\label{sec:results01}A minimal working example}
To demonstrate how to initialize and use the GNLS data structure for performing pulse propagation in terms of Eq.~(\ref{eq:GNLS}), we provide a minimal example in code-listing \ref{code:01}. The provided code initializes the GNLS data structure in lines 10--28,
sets up the input pulse in line 30, retrieves an instance of input pulse noise in line 31,
and propagates the initial field for $10\,\mathrm{cm}$ using the fourth-order Runge Kutta in the interaction picture (RK4IP) method \cite{Hult:JLT:2007}. This type of algorithmic approach is also referred to as integrating factor method \cite{Trefethen:BOOK:2000}, or linearly exact Runge Kutta method \cite{Archilla:JCP:1995}. Pulse propagation is performed in lines 35--42 using a fixed stepsize of $10\,\mathrm{\mu m}$, and using a temporal domain of extent $T=7\,\mathrm{ps}$ with $N=2^{13}$ grid points. 
The script terminates in just under $15\,\mathrm{sec.}$ (Apple M1 chip @ 3.2 GHz) and reproduces Figs.~\ref{fig:01}(a,b) [cf.\ Figs.~18(a,b) of Ref.~\cite{Dudley:RMP:2009}, Fig.~2(a) of Ref.~\cite{Hult:JLT:2007}, Fig.~1(a) of Ref.~\cite{Rieznik:IEEEPJ:2012}, and Fig.~2(b,c) of Ref.~\cite{Melchert:CPC:2022}].
The script shown in listing \ref{code:01} is located in project folder \verb!numExp01! provided with the code \cite{GNLStools:GitHub:2022}.

\subsection{\label{sec:results02}Using GNLS with {\tt py-fmas}}

\begin{figure}[t!]
\includegraphics[width=\linewidth]{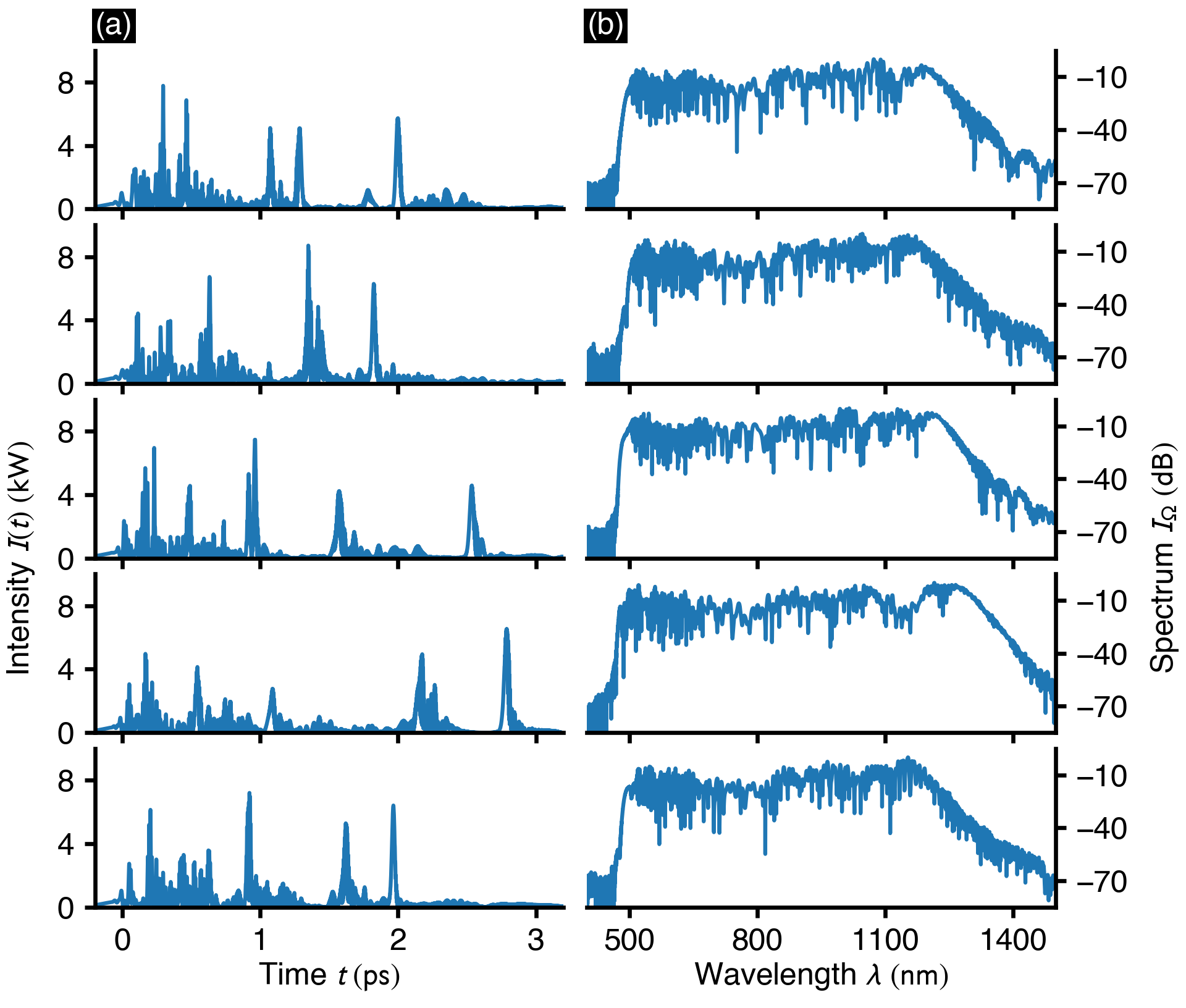}
\caption{Shot-to-shot fluctuations of
(a) instantaneous power, and, (b) corresponding spectra, 
for input pulses with $t_0=85.0\,\mathrm{fs}$ duration input pulses and $10\,\mathrm{cm}$ of propagation (cf.\ Fig.~18 of Ref.~\cite{Dudley:RMP:2009}). 
\label{fig:02}}
\end{figure}

\begin{figure}[t!]
\includegraphics[width=\linewidth]{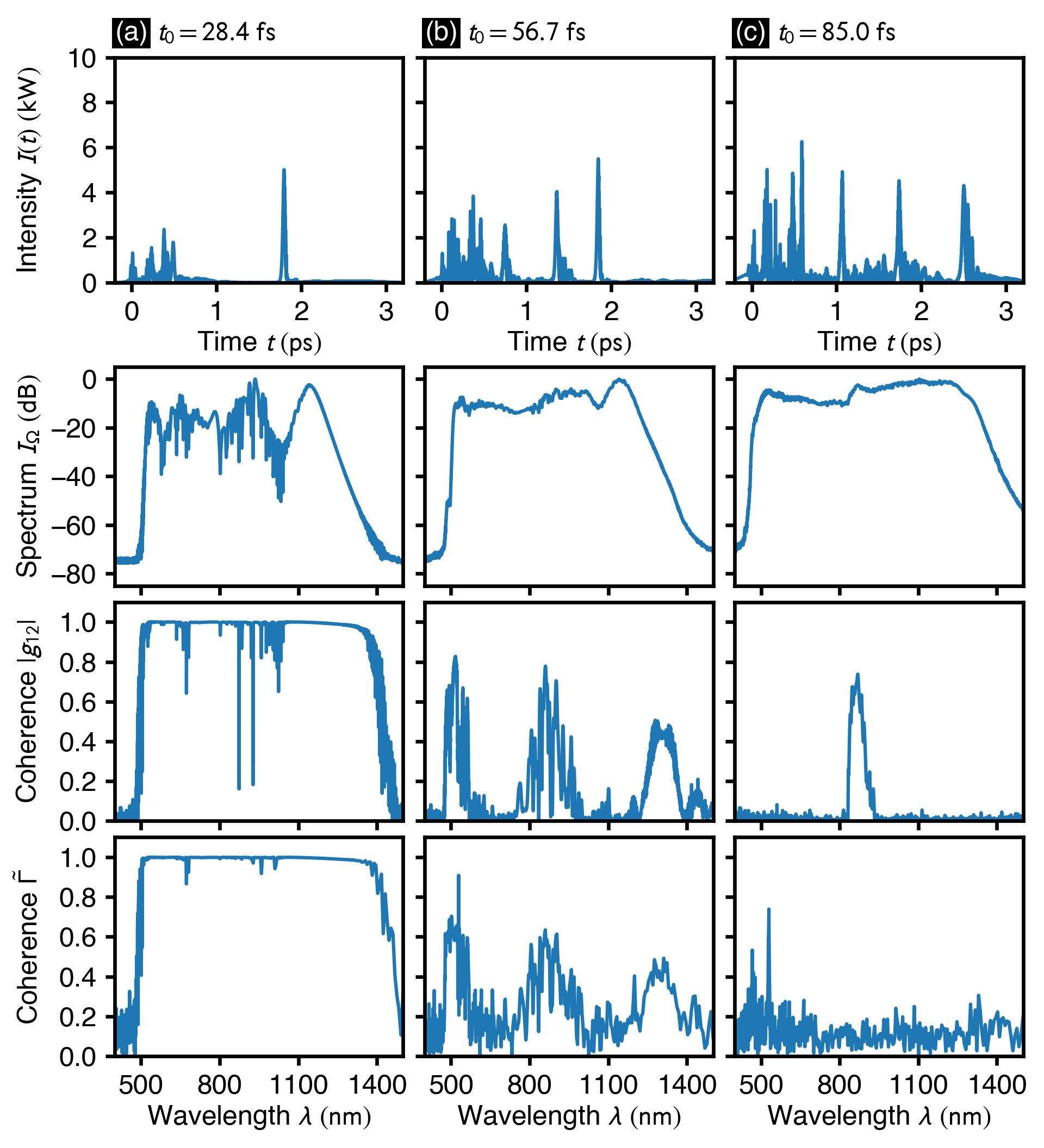}
\caption{Coherence properties of supercontinuum spectra after $10\,\mathrm{cm}$ of propagation (cf.\ Fig.~19 of Ref.~\cite{Dudley:RMP:2009}).
(a) from top to bottom: exemplary single-shot output intensity, ensemble averaged spectrum, and degrees of coherence $|g_{12}|$ [Eq.~(\ref{eq:g})], and $\tilde{\Gamma}$ [Eq.~(\ref{eq:IPC})], as function of wavelength $\lambda = 2 \pi c/(\omega_0+\Omega)$. 
For the calculation of the modified intrapulse coherence $\tilde{\Gamma}(\Omega_1,\Omega_2)$, $\Omega_1=\Omega$, and $\Omega_2=-0.96\,\mathrm{rad/fs}$, corresponding to $\lambda_1=\lambda$, and $\lambda_2=1.45\,\mathrm{nm}$.
Duration of the input pulses is $t_0=28.4\,\mathrm{fs}$.
(b) same for $t_0=56.7\,\mathrm{fs}$, and 
(c) same for $t_0=85.0\,\mathrm{fs}$. 
\label{fig:03}}
\end{figure}

\paragraph{Integration with {\tt py-fmas}}
Figure~\ref{fig:01} is produced by using the {\tt GNLS} data structure in conjunction with {\tt py-fmas} \cite{Melchert:CPC:2022}, a Python package for the numerical simulation of the $z$-propagation dynamics of ultrashort optical pulses in terms of the analytic signal of the optical field. 
{\tt GNLStools} can be used as an elaboration module, providing an envelope based model which integrates well with the propagation algorithms provided by {\tt py-fmas}. 
Specifically, pulse propagation in Figs.~\ref{fig:01}-\ref{fig:03} is performed by the variable stepsize conservation quantity error method  \cite{Heidt:JLT:2009,Rieznik:IEEEPJ:2012,Melchert:CPC:2022}, using Eq.~(\ref{eq:CN}) to guide step-size selection.
Figure~\ref{fig:01}(e) shows the numerical error accumlated for the photon number [Eq.~(\ref{eq:CN})] and energy [Eq.~(\ref{eq:E})], showing that $C_{\rm{Ph}}$ is conserved up to order $10^{-8}$, while $E$ is not conserved \cite{Blow:JQE:1989}. 

Considering $C_{\rm{Ph}}$, the feature at $z\approx 1~\mathrm{cm}$ indicates the onset of soliton-fission (close-up view in Fig.~\ref{fig:01}c), and the change of the trend at $z\approx 7~\mathrm{cm}$ indicates the onset of the interaction of a soliton and a dispersive wave which mainly determines the interaction dynamics and generation of supercontinuum spectra.
A small project workflow with a driver script that performs the simulation, and a postprocessing script that generates Fig.~\ref{fig:01}, are located in project folder \verb!numExp02! provided with the code \cite{GNLStools:GitHub:2022}.

\begin{lstlisting}[float=*,floatplacement=H, numbers=left, 
captionpos=t, frame=lines,stepnumber=1,numbers=left, numbersep=5pt, xleftmargin=\parindent,
language=Python, 
caption={Python code for solving the GNLS (\ref{eq:GNLS}) using the fourth-order Runge Kutta in the interaction picture method \cite{Hult:JLT:2007}.}, label=code:01]
import numpy as np
import matplotlib.pyplot as plt
from GNLStools import GNLS, noise_model_01

# -- SET COMPUTATIONAL GRID
z, dz = np.linspace(0, 0.1e6, 10000, retstep=True)
t = np.linspace(-3500, 3500, 2**13, endpoint=False)
w = np.fft.fftfreq(t.size, d=t[1]-t[0])*2*np.pi
# -- INSTANTIATE GENERALIZED NONLINEAR SCHROEDINGER EQUATION 
gnls = GNLS(
    w,               # (rad/fs)
    beta_n = [
        -1.1830e-2,  # (fs^2/micron) beta_2
        8.1038e-2,   # (fs^3/micron) beta_3
        -0.95205e-1, # (fs^4/micron) beta_4
        2.0737e-1,   # (fs^5/micron) beta_5
        -5.3943e-1,  # (fs^6/micron) beta_6
        1.3486,      # (fs^7/micron) beta_7
        -2.5495,     # (fs^8/micron) beta_8
        3.0524,      # (fs^9/micron) beta_9
        -1.7140,     # (fs^10/micron) beta_10
        ],
    gamma=0.11e-6,   # (1/W/micron)
    w0= 2.2559,      # (rad/fs)
    fR = 0.18,       # (-)
    tau1 = 12.2,     # (fs)
    tau2 = 32.0      # (fs)
    )
# -- SPECIFY INITIAL PULSE
ut = np.sqrt(1e4)/np.cosh(t/28.4)
dut = noise_model_01(t, 2.2559, 1)
uw = np.fft.ifft(ut + dut)

# -- RK4IP PULSE PROPAGATION
P = np.exp(gnls.Lw*dz/2)
for n in range(1,z.size):
    uw_I = P*uw
    k1 = P*gnls.Nw(uw)*dz
    k2 = gnls.Nw(uw_I + k1/2)*dz
    k3 = gnls.Nw(uw_I + k2/2)*dz
    k4 = gnls.Nw(P*uw_I + k3)*dz
    uw = P*(uw_I + k1/6 + k2/3 + k3/3) + k4/6

# -- PLOT RESULTS
fig, (ax1, ax2) = plt.subplots(1, 2, figsize=(8, 3))
I = np.abs(np.fft.fft(uw))**2
ax1.plot(t, I*1e-3)
ax1.set_xlim(-200,3200); ax1.set_xlabel(r"Time $t$ (fs)")
ax1.set_ylim(0,6); ax1.set_ylabel(r"Intensity $|u|^2$ (kW)")
Iw = np.abs(uw)**2
ax2.plot(2*np.pi*0.29979/(w+2.2559), 10*np.log10(Iw/np.max(Iw)))
ax2.set_xlim(0.45,1.4); ax2.set_xlabel(r"Wavelength $\lambda$ (micron)")
ax2.set_ylim(-60,0); ax2.set_ylabel(r"Spectrum $|u_\lambda|^2$ (dB)")
fig.tight_layout(); plt.show()
\end{lstlisting}

\paragraph{Shot-to-shot fluctuations}
Figure~\ref{fig:02} demonstrates shot-to-shot variations in pulse intensity and spectrum after $10\,\mathrm{cm}$ of propagation, arising from the inclusion of input pulse noise for an input pulse with duration $t_0=85.1\,\mathrm{fs}$. The noise is generated via noise model 1, i.e.\ by direct sampling in the time-domain.
A small project workflow with a driver script that performs the simulation, and a postprocessing script that generates Fig.~\ref{fig:02}, are located in project folder \verb!numExp03_noise_model_01!, provided with the code \cite{GNLStools:GitHub:2022}.

\paragraph{Coherence properties}
The coherence properties of input pulses of different duration, obtained by performing ensemble averages over $200$ independent simulation runs with different noise seeds, are shown in Fig.~\ref{fig:03}. 
Figure~\ref{fig:03} reproduces Fig.~19 of Ref.~\cite{Dudley:RMP:2009}, wherein a detailed discussion of the effects of input pulse noise and the interpretation of the coherence can be found. 
In addition, in Fig.~\ref{fig:03} we also show the modified intrapulse coherence [Eq.~\ref{eq:IPC}] discussed in Ref.~\cite{Raabe:PRL:2017}.
A postprocessing script that generates Fig.~\ref{fig:03} is located in folder \verb!numExp03_noise_model_01! provided with the code \cite{TBW}.
From our experience, the slight correlations introduced by the Fourier method based noise models, see Sect.~\ref{sec:noise_models} and Fig.~\ref{fig:00}, do not affect the coherence properties of the obtained supercontinuum spectra.

\section{Impact and conclusions \label{sec:conclusions}}

The presented {\tt GNLStools} comprise a data structure implementing the generalized nonlinear Schrödinger equation, two commonly adopted models of quantum noise and a further noise model based on a classical analog of the zero-point field of QFT, and functions for assessing the coherence properties of simulated spectra.
It provides all features required for studying basic phenomena supported by the GNLS in the presence of input pulse shot noise.
%
%
The provided software can be used on its own, as demonstrated in Sec.\ \ref{sec:results01}, or as elaboration module for use with the propagation algorithms provided by the {\tt py-fmas} package \cite{Melchert:CPC:2022}, as demonstrated in Sec.\ \ref{sec:results02}.

With the provided software we hope to shed some more light on the implementation of quantum noise models for use in pulse propagation studies, an issue that is often not thoroughly presented in the scientific literature. We hope to directly benefit students and researchers alike, which are new to the field of nonlinear optics and seek a tutorial-type introduction on how to perform pulse propagation simulations including quantum noise, and assess the coherence properties of the resulting spectra.
The minimal example in Sec.\ \ref{sec:results01} can very well serve as classroom code or as starting point for seminar projects in computation oriented courses, aimed at demonstrating algorithmic approaches that go beyond simple split-step Fourier methods commonly used for solving nonlinear Schrödinger-type equations \cite{Agrawal:BOOK:2019,Taha:JCP:1984,DeVries:AIP:1987}.

Finally, we would like to reference original research applying the presented GNLS tools. 
In Ref.~\cite{Melchert:NCP:2020} we used the presented software to demonstrate an efficient all-optical switching scheme, based on controlling the features of soliton fission induced supercontinuum spectra using a time-delayed dispersive wave.

\section*{Declaration of competing interests}
The authors confirm that there are no known conflicts of interest associated
with this publication.

\section*{Acknowledgements}

We acknowledge support from the Deutsche Forschungsgemeinschaft  (DFG) under
Germany’s Excellence Strategy within the Cluster of Excellence PhoenixD
(Photonics, Optics, and Engineering – Innovation Across Disciplines) (EXC 2122,
projectID 390833453).


\addcontentsline{toc}{section}{References}
\bibliography{references}

\begin{thebibliography}{10}
\bibitem{note:FT}
On the discrete computational grid we identify $\mathsf{F}$ with {\tt ifft},
  and $\mathsf{F}^{-1}$ with {\tt fft}. Consequently, the spectral envelope in
  the provided implementation differs from the one defined by
  Eq.~(\ref{eq:FT_FT}) by an unimportant phase factor.
\end{thebibliography}

\end{document}